\documentclass[11pt]{amsart}
\usepackage{graphicx} 
\usepackage[utf8]{inputenc}
\usepackage{tikz-cd}
\usepackage{tikz}
\usepackage[left=3.5cm,right=3.5cm,top=3.5cm,bottom=3.5cm]{geometry}
\usepackage{comment}
\usepackage{amsfonts}
\usepackage{amsmath}
\usepackage{amssymb}
\usepackage{amsthm}
\usepackage{mathrsfs}
\newtheorem{theorem}{Theorem}
\newtheorem{lemma}{Lemma}

\newtheorem{corollary}{Corollary}
\newtheorem{definition}{Definition}
\theoremstyle{remark}

\newtheorem{example}{Example}

\usepackage{stmaryrd}

\usepackage[svgnames]{xcolor}
\usepackage[colorlinks=true, linkcolor=Maroon, urlcolor=Maroon, citecolor=Maroon, filecolor=Maroon]{hyperref}

\usepackage[
backend=biber,
style=numeric,
sorting=ynt
]{biblatex}
\usepackage{hyperref}
\begin{document}
\title{Risk Measure Duality Without Structure}
\author{Vasily Melnikov}
\date{August 2026}
\address{University of Alberta, Edmonton, AB, Canada.}

\begin{abstract}
     We study risk measures on vector spaces of random variables which a priori have little structure, such as spaces lacking law invariance or a lattice structure. Ensuring the existence of a tractable dual representation (one which does not contain non-sigma-additive measures) is one of the main problems in risk measure theory, and we address it under minimal conditions. The existence of a tractable dual representation is shown to be equivalent to a Fatou-like property when the domain of the risk measure satisfies a topological regularity condition. Without the topological regularity condition, the Fatou property implies the existence of a tractable dual representation whenever the risk measure is viewed with constraints. We also present counterexamples demonstrating the sharpness of the assumptions made.
\end{abstract}

\maketitle

\section{Introduction}\label{sec:intro}
The concept of a coherent risk measure was introduced by Artzner, Delbaen, Eber, and Heath \cite{coherent}, later extended to the more general theory of convex risk measures. A natural and important question is the existence of a ``robust dual representation'' for such functionals, the study of which was initiated by Delbaen \cite{delrisk} for functionals on the space $L^{\infty}$ of bounded random variables. Dual representations of risk measures are an important tool for a wide range of problems involving risk measures, including risk sharing (see \cite{law-inv-util-conv-dual,heath-ku}), optimization of risk measures (see \cite{continuity-of-monotone-functionals}), and capital allocation (see \cite{cap-alloc-sys}).
\par
A significant literature has studied dual representation results on spaces of unbounded random variables. For example, results are known for Orlicz spaces (see \cite{ngaomf} and \cite{extorl}), Orlicz hearts (see \cite{banlattrisk} and \cite{studiagao}), and the space $L^{0}$ of all random variables (see \cite{duall0}). We take a different approach, focusing less on the specific structure of a given space $\mathcal{X}$ on which a risk measure is defined, and more on the minimum structure necessary to extend dual representation results to $\mathcal{X}$ and the relation between the Fatou property and dual representations. Such an approach is useful because risk measures have been considered on a diversity of spaces, even those that are not classical in the theory of Banach function spaces. For example, in the literature there is a wealth of constructions for a suitable risk measure $\varrho$ that defines a maximal domain for $\varrho$, whose structure may differ from the case of $L^{p}$ and Orlicz spaces (see, e.g., \cite{natural-ban-space-risk,lebext}).
\par
Our approach is thus abstract, beginning with a vector space $\mathcal{X}$ of random variables. Some structure is assumed—for example, $\mathcal{X}$ must be built up from bounded parts, in a certain precise sense, allowing one to endow $\mathcal{X}$ with a norm. The assumptions are tailored to ensure the minimum structure necessary to formulate the concept of dual representations and the Fatou property, but little more. There is significant precedent for such an abstract approach, but also some important differences with the previous literature on this topic. For example, we do not assume law invariance of the domain (cf. \cite{belllaw}), as this is not strictly necessary in order to define the Fatou property. Law invariance is assumed in many risk measures contexts, so its failure in a particular situation often limits the available literature. However, our approach is also more grounded and specific than other abstract approaches, which sometimes do not assume a fixed probabilistic structure at all, formulating results in the context of Fréchet lattices (see \cite{frechetrisk}), ordered topological vector spaces (see \cite{farkas-koch-medin-mun,put-order}), model uncertainty (see \cite{reversefelix}),\footnote{Of course, the usual approach to even non-dominated model uncertainty does assume some probabilistic structure, but the point is that this cannot come from a fixed probability space. But see Cohen's notion of the Hahn property \cite{hahn-cohen}, showing non-dominated model uncertainty can sometimes be captured by a fixed strictly localizable measure space.} or $C_{b}$ spaces (see \cite{nendel_cb}). In such a case, the Fatou property is usually formulated using the concept of order convergence (see, e.g., \cite[\S5.4.F2]{reversefelix}), which is not in general reducible to sequences as in the classical Fatou property.\footnote{For sequences, order convergence is equivalent, modulo a subsequence, to convergence in probability conjoined with order-boundedness (see, e.g., \cite[Remark 3.4]{ces}), and thus the difficulties of order convergence stem from having to consider general nets, rather than just sequences.} In contrast, we formulate the Fatou property using norm-bounded sequences converging in probability, a construction less free of dependence on the underlying probabilistic model than order convergence, but also often easier to verify and work with in practice.
\par
Under our framework for the domain space $\mathcal{X}$, it is possible to define a dual space $\mathcal{Y}$ for $\mathcal{X}$ in terms of a polar set construction introduced by Brannath and Schachermayer \cite{bipolar}; $\mathcal{Y}$ consists of random variables, and is thus in general different from the Banach space dual, but is not in general unique. As a consequence, it is possible to formulate dual representations for functionals on $\mathcal{X}$ in terms of this dual space, as well as some weak-star-type topologies on $\mathcal{X}$.
\par
Our first major result relates the Fatou property on $\mathcal{X}$ to lower semicontinuity on bounded subsets of $\mathcal{X}$ with respect to a locally convex topology on $\mathcal{X}$. The Fatou property is essentially a lower semicontinuity assumption for the topology of convergence in probability, with the understanding that bona fide lower semicontinuity in this topology is often not possible under non-trivialities.\footnote{A lower semicontinuity property for almost sure convergence was introduced by Wang and Zitikis \cite{wang-zit} and termed prudence.} This generalizes the result that, on $L^{\infty}$, the Fatou property is equivalent to lower $\sigma(L^{\infty},L^{1})$-semicontinuity, connecting the $L^{\infty}$-$L^{1}$ duality to convergence in probability.
\par
The equivalence between the Fatou property and certain kinds of lower semicontinuity allow the deduction of a dual representation result. Dual representations are a useful tool, but their existence may sometimes be subtle and has to be verified or assumed before dual representations can be used. Supposing that a version of the Krein-Šmulian theorem from functional analysis holds, we show every convex functional on $\mathcal{X}$ satisfying the Fatou property has a dual representation. If $\mathcal{Y}$ denotes the dual of the domain space $\mathcal{X}$, the dual representation (if it exists) for a proper convex functional $\varrho:\mathcal{X}\longrightarrow\mathbb{R}\cup\{\infty\}$ takes the form
\begin{equation*}
    \varrho(X)=\sup_{Y\in \mathcal{Y}}\left\{\mathbb{E}(XY)-\alpha(Y)\right\},
\end{equation*}
where $\alpha:\mathcal{Y}\longrightarrow\mathbb{R}\cup\{\infty\}$ denotes a penalty function. Importantly, the dual space $\mathcal{Y}$ consists of random variables, and so the dual representation does not contain any pathological non-$\sigma$-additive elements. Since every finite monotone convex functional on a Banach lattice is norm continuous (see, e.g., \cite{continuity-of-monotone-functionals}), and hence has a dual representation with respect to the norm dual, clarifying whether the dual representation has non-$\sigma$-additive elements is usually the primary question.
\par
In certain cases, however, the validity of the Krein-Šmulian theorem is nontrivial or otherwise not satisfied (see \cite{studiagao}). Where there are such concerns, we show a general dual representation results—under constraints. If a convex non-empty subset $C\subseteq\mathcal{X}$ is closed in probability and bounded in $\mathcal{X}$, then we show any proper convex functional $\varrho:\mathcal{X}\longrightarrow\mathbb{R}\cup\{\infty\}$ takes the form
\begin{equation*}
    \varrho(X)=\sup_{Y\in \mathcal{Y}}\left\{\mathbb{E}(XY)-\alpha(Y)\right\},
\end{equation*}
where $\alpha:\mathcal{Y}\longrightarrow\mathbb{R}\cup\{\infty\}$ denotes a penalty function, whenever $\varrho(X)$ is considered only for $X\in C$.
\par
Our paper is structured as follows. In \S\ref{sec:prelim}, we establish some preliminaries. In \S\ref{sec:model}, we introduce our framework. In \S\ref{sec:lsc}, we apply our framework to questions of lower semicontinuity of convex functionals. In \S\ref{sec:dualrep}, our lower semicontinuity results are applied to the problem of dual representations.
\section{Preliminaries}\label{sec:prelim}
Let $(\Omega,\mathscr{F},\mathbb{P})$ be a probability space; $L^{0}(\mathbb{P})$ denotes the space of equivalence classes (modulo $\mathbb{P}$-a.s. equality) of real-valued random variables, and $L^{0}_{+}(\mathbb{P})$ denotes its positive orthant $\{X\in L^{0}(\mathbb{P}):X\geq0\}$. The space $L^{p}(\mathbb{P})$, where $1\leq p<\infty$, consists of all $X\in L^{0}(\mathbb{P})$ with $\mathbb{E}^{\mathbb{P}}(\vert{X}\vert^{p})<\infty$. The space $L^{\infty}(\mathbb{P})$ consists of all essentially bounded $X\in L^{0}(\mathbb{P})$. The unit ball of $L^{p}(\mathbb{P})$, for $1\leq p\leq\infty$, is denoted $B_{L^{p}}$.
\par
Let $K\subseteq L^{0}(\mathbb{P})$. $K$ is said to be solid if $\vert{X}\vert\leq\vert{Y}\vert$ and $Y\in K$ implies $X\in K$. $K$ is said to be bounded in probability if, for each $\varepsilon>0$, there exists $\delta>0$ such that
\begin{equation*}
    \sup_{X\in K}\mathbb{P}\left(\left\{\vert{X}\vert\geq\delta\right\}\right)\leq\varepsilon.
\end{equation*}
Put differently, $K$ is bounded in probability if the members of $K$ are pointwise bounded up to some small probability. The polar of $K$ is the set $K^{\circ}\subseteq L^{0}(\mathbb{P})$ defined by
\begin{equation*}
    K^{\circ}=\left\{Y\in L^{0}(\mathbb{P}):\sup_{X\in K}\mathbb{E}^{\mathbb{P}}\left(\vert{XY}\vert \right)\leq1\right\}.
\end{equation*}
The bipolar of $K$ is $K^{\circ\circ}=(K^{\circ})^{\circ}$.
\par
Given $A\subseteq B$, we will denote by $\mathbf{1}_{A}$ the indicator of $A$, and by $\mathbb{I}_{A}=\infty\mathbf{1}_{B\setminus A}$ the characteristic functional of $A$.
\par
If $(E,\tau)$ is a topological vector space, then $(E,\tau)^{\ast}$ will denote the topological dual of $(E,\tau)$. Given a duality pairing $\langle{\cdot,\cdot}\rangle:E\times F\longrightarrow\mathbb{R}$, we will denote by $\sigma(E,F)$ the weak topology associated to that duality. The topology $\sigma(E,F)$ is locally convex, being generated by the seminorms $\vert{\langle{\cdot,f}\rangle}\vert$ for $f\in F$.
\section{The Model}\label{sec:model}
We formulate relatively minimal assumptions on a vector space $\mathcal{X}$ of random variables so that notions important to risk measures—for example, the Fatou property, dual spaces, and dual representations—can be defined on $\mathcal{X}$.
\subsection{Boundedness}\label{subsec:bounded}
The first important assumption on $\mathcal{X}$ is that it admits a notion of boundedness. This is important not only for technical concerns, but also economic reasons.
\par
Consider, for example, the martingale strategy in the St. Petersburg paradox of Bernoulli \cite{bern}. At each step, the strategy on average yields $\$0$. However, when all steps are viewed together, the limiting net profit is a sure $\$1$. Plainly, this means the Fatou property of the expected value cannot be understood to extend to \textit{unbounded} sequences of random variables, as the final expected value is not majorized, even in a limiting sense, by the expectation of each step of the strategy. From an economic point of view, this corresponds to the understanding that if the risk of some strategy is unbounded, the interim net profits can be very negative, yet in the end always become positive—yielding useless extrapolations of the expected profit and risk of the end value from the interim values. Excluding such pathological cases from the definition of the Fatou property is necessary to ensure the Fatou property is not too narrow, and this is exactly the sort of demarcation a notion of boundedness makes.\footnote{An example of such a demarcation in a related context, that of continuous time arbitrage theory, is the stipulation that a sequence of strategies is only said to violate no arbitrage conditions if the payoffs converge to a non-negative and non-zero payoff, and the interim losses are bounded below in $L^{\infty}$ (see \cite{del-sch}).}
\par
To define a notion of boundedness, let $K\subseteq \mathcal{X}$ be absolutely convex, closed and bounded in probability, and absorbing—meaning $\mathcal{X}$ is ``generated'' by $K$, in the sense that every $X\in \mathcal{X}$ is a member of $\lambda K$ for some $\lambda>0$. $K$ yields a norm, and hence a notion of boundedness, via the Minkowski functional $p_{K}$:
\begin{equation*}
    p_{K}(X)=\inf\left\{r>0:X\in rK\right\}.
\end{equation*}
To make this conception of boundedness more explicit, a set $C\subseteq\mathcal{X}$ is said to be \textit{$K$-bounded} if $\sup_{X\in C}p_{K}(X)<\infty$. By an abuse of notation, the extra data $K$ together with $\mathcal{X}$ will be denoted $\mathcal{X}$.
\subsection{Fatou Property}
With boundedness in $\mathcal{X}$ defined by the extra data of $K$, we may now formulate a version of the Fatou property for functionals $\varrho:\mathcal{X}\longrightarrow\mathbb{R}\cup\{\infty\}$. As noted in \S\ref{subsec:bounded}, it is important to restrict the Fatou property's scope using boundedness, and hence the Fatou property we define in this section uses a notion of boundedness in its definition.
\begin{definition}\label{def:fatou}
    A functional $\varrho:\mathcal{X}\longrightarrow\mathbb{R}\cup\{\infty\}$ is said to have the $K$-Fatou property if
    \begin{equation*}
        \varrho(X)\leq\liminf_{n\to\infty}\varrho(X^{n})
    \end{equation*}
    for every $K$-bounded sequence $(X^{n})_{n=1}^{\infty}\subseteq\mathcal{X}$ converging in probability to some $X\in \mathcal{X}$.
\end{definition}
For $\varrho$ to satisfy Definition \ref{def:fatou}, it is necessary that, for a sufficiently regular approximation of a random payoff $X$, the corresponding approximation of $\varrho(X)$ provides at least an upper bound for the actual value of $\varrho(X)$.
\par
Definition \ref{def:fatou}, with its emphasis on a notion of boundedness defined by a norm, does differ slightly from some other Fatou properties in the literature. Weaker definitions might only require a Fatou-type condition on convergent order bounded sequences (see, e.g., \cite{frechetrisk}) or order convergent sequences (see, e.g., \cite{reversefelix}). But Definition \ref{def:fatou} is not an egregious loss of generality, for order boundedness and order convergence become unnatural without a vector lattice structure, which we do not assume, and in certain situations the various formulations are already known to be equivalent (such as Orlicz spaces $L^{\Phi}$, unless both $\Phi$ and $\Phi^{\ast}$ fail the $\Delta_{2}$-condition; see \cite{studiagao} on this point).
\subsection{Duality}
Given $\mathcal{X}$ and $K$, it is possible to define a set of ``dual random variables'' or ``prices'' on $\mathcal{X}$. More precisely, consider the \textit{polar set} $K^{\circ}$ of $K$:
\begin{equation*}
    K^{\circ}=\left\{Y\in L^{0}(\mathbb{P}):\sup_{X\in K}\mathbb{E}^{\mathbb{P}}\left(\vert{XY}\vert\right)\leq1\right\}.
\end{equation*}
Then every $Y\in K^{\circ}$ defines a linear pricing rule on $\mathcal{X}$, defined by $X\longmapsto\mathbb{E}^{\mathbb{P}}(XY)$. Similarly, every $Y\in\mathrm{span}(K^{\circ})$ defines a linear pricing rule on $\mathcal{X}$ in an exactly analogous manner. Thus, $\mathrm{span}(K^{\circ})$ can be viewed as a dual space to $\mathcal{X}$, and we denote $\mathrm{span}(K^{\circ})$ by $\mathcal{X}'_{K}$. Of course, any subspace $\mathcal{Y}\subseteq\mathcal{X}'_{K}$ also can be viewed as a dual space for $\mathcal{X}$.
\par
In general, even if $K$ is bounded in probability, it is possible for $K^{\circ}=\{0\}$ (an example is given in Appendix \ref{app:trivial-dual}). However, $K^{\circ}$ is sufficiently rich—for example, separating the points of $\mathcal{X}$—for many purposes under rather weak conditions, like the assumption $K\subseteq B_{L^{1}}$, in which case $B_{L^{\infty}}\subseteq K^{\circ}$. The assumption $K\subseteq B_{L^{1}}$ often allows the reduction of problems about dual representations to the corresponding questions about $L^{\infty}$ and $L^{1}$ (see \cite{extl1,owari-conv}), and we apply a similar observation in \S\ref{subsec:lemmas} and \S\ref{subsec:prf}.
\subsection{Topological Structures}
Being a space of random variables, convergence in probability of random variables in $\mathcal{X}$ is well-defined—but the resulting topology is often not useful. Compared to a norm topology, convergence in probability is too coarse, supports few continuous linear mappings, and often fails to be locally convex (see, e.g., \cite{kardlocconv,poslocconv}). The literature has thus experimented with substitutes for convergence in probability.
\par
For example, Delbaen \cite{delrisk}, and later Gao and Xanthos \cite{ngaomf}, use the weak star topology to establish dual representation theorems. In the case of $L^{\infty}(\mathbb{P})$, some of the important properties of this topology, denoted $\sigma(L^{\infty},L^{1})$, are the following.
\begin{enumerate}
    \item\label{item:compact} The unit ball $B_{L^{\infty}}$ of the $L^{\infty}(\mathbb{P})$-norm is $\sigma(L^{\infty},L^{1})$-compact.
    \item\label{item:solid-dual} The dual of $(L^{\infty}(\mathbb{P}),\sigma(L^{\infty},L^{1}))$ is $L^{1}(\mathbb{P})$, which is a solid subspace of $L^{0}(\mathbb{P})$ containing a strictly positive element.
    \item\label{item:ks} The Krein-Šmulian theorem holds: if $C\subseteq L^{\infty}(\mathbb{P})$ is convex, and $C\cap\lambda B_{L^{\infty}}$ is $\sigma(L^{\infty},L^{1})$-closed for each $\lambda\geq0$, then $C$ is $\sigma(L^{\infty},L^{1})$-closed.
\end{enumerate}
In an analogy to properties \ref{item:compact}-\ref{item:solid-dual}, we define the notion of a $K$-equicontinuous topology, and the corresponding variant of \ref{item:ks}, the Krein-Šmulian property.
\begin{definition}\label{def:equitop}
    A topology $\tau$ on $\mathcal{X}$ is said to be $K$-equicontinuous if $K$ is $\tau$-compact, and $\tau=\sigma(\mathcal{X},\mathcal{Y})$, where the linear subspace $\mathcal{Y}\subseteq\mathcal{X}'_{K}$ is solid and contains a strictly positive element. In this case, $\mathcal{Y}$ is said to induce $\tau$.
\end{definition}
In the context of our later results, the assumptions made in the definition of a $K$-equicontinuous topology are sharp (see Appendix \ref{app:necessary-compact}).
\begin{definition}
    A topology $\tau$ on $\mathcal{X}$ is said to have the $K$-Krein-Šmulian property if the $\tau$-closedness of a convex $C\subseteq\mathcal{X}$ is equivalent to the $\tau$-closedness, for all $\lambda\geq0$, of $C\cap \lambda K$.
\end{definition}
The existence of a $K$-equicontinuous topology on $\mathcal{X}$ is a relatively weak assumption. For example, $\tau$-compactness of $K$ is later shown to be related to uniform integrability of $K$, and the existence of a strictly positive element in $\mathcal{X}'_{K}$ (of which $\mathcal{Y}$ is a subspace) is equivalent to boundedness of $K$ in $L^{1}(\mathbb{Q})$ for some $\mathbb{Q}\sim\mathbb{P}$. On the other hand, the Krein-Šmulian property is more difficult to assume (see \cite{studiagao}), but this is remedied by the non-necessity of the Krein-Šmulian property for many of the results in the sequel.
\par
If $\tau$ is $K$-equicontinuous, the $K$-Krein-Šmulian property will be referred to as the Krein-Šmulian property. $K$-equicontinuous topologies need not have the Krein-Šmulian property in general, as the following example shows.
\begin{example}
    Consider $\mathcal{X}=L^{2}(\mathbb{P})$, $K=B_{L^{2}}$, and $\mathcal{Y}=L^{\infty}(\mathbb{P})$. Then $\tau=\sigma(\mathcal{X},\mathcal{Y})$ is $K$-equicontinuous, but fails the Krein-Šmulian property whenever $\mathbb{P}$ is non-atomic: if $C=\{X\in\mathcal{X}:\mathbb{E}^{\mathbb{P}}(XY)=0\}$ for some fixed $Y\in L^{2}(\mathbb{P})\setminus L^{\infty}(\mathbb{P})$, then $C$ is convex and $C\cap\lambda K$ is $\tau$-closed for each $\lambda\geq0$, but $C$ is not $\tau$-closed.
    \qed
\end{example}
\section{Lower Semicontinuity}\label{sec:lsc}
For $K$-equicontinuous topologies $\tau$, we may relate the Fatou property of convex proper functionals $\varrho:\mathcal{X}\longrightarrow\mathbb{R}\cup\{\infty\}$ with lower $\tau$-semicontinuity, albeit only ``locally''. We elaborate on this below:
\begin{theorem}\label{thm:semicont}
    Let $\varrho:\mathcal{X}\longrightarrow\mathbb{R}\cup\{\infty\}$ be a convex and proper functional. Then, for any $K$-equicontinuous topology $\tau$, the following are equivalent.
    \begin{enumerate}
        \item\label{it:lsc} $\{\varrho\leq\lambda\}\cap \lambda' K$ is $\tau$-closed for any $\lambda\in\mathbb{R}$ and any $\lambda'\geq0$.
        \item\label{it:fatou} $\varrho$ has the $K$-Fatou property.
    \end{enumerate}
\end{theorem}
Under the Krein-Šmulian property, closedness of the ``local'' sublevel sets from Theorem \ref{thm:semicont}, $\{\varrho\leq\lambda\}\cap\lambda' K$ for $\lambda\in\mathbb{R}$ and $\lambda'\geq0$, can be related to closedness of the ``global'' sublevel sets $\{\varrho\leq\lambda\}$. Hence, the relation established between the former and the Fatou property by Theorem \ref{thm:semicont} extends to an equivalence between the Fatou property and lower semicontinuity whenever the Krein-Šmulian property holds.
\begin{corollary}\label{corr:semicont-ks}
    Let $\varrho:\mathcal{X}\longrightarrow\mathbb{R}\cup\{\infty\}$ be a convex and proper functional, and let $\tau$ be a $K$-equicontinuous topology with the Krein-Šmulian property. Then the following are equivalent.
    \begin{enumerate}
        \item\label{it:fatou-corr} $\varrho$ has the $K$-Fatou property.
        \item\label{it:lsc-corr} $\varrho$ is lower $\tau$-semicontinuous.
    \end{enumerate}
\end{corollary}
\begin{proof}
    If $\varrho$ has the $K$-Fatou property, then $\{\varrho\leq\lambda\}\cap\lambda' K$ is $\tau$-closed for each $\lambda\in\mathbb{R}$ and $\lambda'\geq0$ by Theorem \ref{thm:semicont}. Since $\{\varrho\leq\lambda\}$ is convex, the Krein-Šmulian property implies $\{\varrho\leq\lambda\}$ is $\tau$-closed. Since $\lambda\in\mathbb{R}$ was arbitrary, $\varrho$ is lower $\tau$-semicontinuous.
    \par
    Conversely, if $\varrho$ is lower $\tau$-semicontinuous, then $\{\varrho\leq\lambda\}\cap\lambda' K$ is $\tau$-closed for each $\lambda\in\mathbb{R}$ and $\lambda'\geq0$, and Theorem \ref{thm:semicont} applies to show $\varrho$ has the $K$-Fatou property.
\end{proof}
The Krein-Šmulian property is necessary for the equivalence in Corollary \ref{corr:semicont-ks} to hold. Indeed, if $C\subseteq\mathcal{X}$ is convex, non-empty, and $C\cap\lambda K$ is $\tau$-closed for each $\lambda\geq0$, but $C$ is not $\tau$-closed, the functional $\varrho=\mathbb{I}_{C}$ is convex, proper, and has the $K$-Fatou property (by virtue of Theorem \ref{thm:semicont}), but cannot be lower $\tau$-semicontinuous. The existence of such a set $C$ is equivalent to the negation of the Krein-Šmulian property.
\subsection{Lemmata}\label{subsec:lemmas}
Here, we deal primarily with the relations between equicontinuous $K$-topologies and uniform integrability. Since the choice of reference measure $\mathbb{P}$ is essentially arbitrary to our framework, a modification of uniform integrability, which implicitly depends on a measure, is required. Such a notion is provided by weak compactizability, introduced by Kardaras (see \cite[Remark 2.6]{kardlocconv}).
\begin{definition}
    A subset $C\subseteq L^{0}(\mathbb{P})$ is said to be weakly compactizable if there exists an equivalent probability measure $\mathbb{Q}\sim\mathbb{P}$ such that $C$ is uniformly $\mathbb{Q}$-integrable.
\end{definition}
The terminology above is justified by the Dunford-Pettis theorem: $C$ is weakly compactizable iff there exists an equivalent probability measure $\mathbb{Q}\sim\mathbb{P}$ such that $C\subseteq L^{1}(\mathbb{Q})$ and $C$ is relatively $\sigma(L^{1}(\mathbb{Q}),L^{\infty})$-compact. 
\par
Whether an equicontinuous $K$-topology exists is characterized by weak compactizability of $K$, as shown by Lemma \ref{lem:weakcompacttopo} below.
\begin{lemma}\label{lem:weakcompacttopo}
$\mathcal{X}$ admits an equicontinuous $K$-topology if, and only if, $K$ is weakly compactizable. In particular, if there is an equicontinuous $K$-topology, there exists a probability measure $\mathbb{Q}\sim\mathbb{P}$ such that $K$ is uniformly $\mathbb{Q}$-integrable.
\end{lemma}
\begin{proof}
Suppose there exists $\mathbb{Q}\sim\mathbb{P}$ such that $K$ is uniformly $\mathbb{Q}$-integrable; we claim that the $\sigma(L^{1}(\mathbb{Q}),L^{\infty})$-subspace topology $\tau$ on $\mathcal{X}$ is an equicontinuous $K$-topology, which would prove the forward implication. The Dunford-Pettis theorem implies $K$ is relatively $\tau$-compact, so it suffices to show that $K$ is closed in $L^{1}(\mathbb{Q})$—an easy consequence of Markov's inequality and closedness in probability of $K$.
\par
We now prove the converse. Suppose $\tau$ is a $K$-equicontinuous topology, induced by some $\mathcal{Y}\subseteq\mathcal{X}'_{K}$. By definition, there must exist a strictly positive $Y\in\mathcal{Y}$. Define an equivalent probability measure $\mathbb{Q}\sim\mathbb{P}$ by its Radon-Nikodým derivative:
\begin{equation*}
    \frac{d\mathbb{Q}}{d\mathbb{P}}=\frac{Y\wedge1}{\mathbb{E}^{\mathbb{P}}\left(Y\wedge1\right)}.
\end{equation*}
We claim that $K$ is uniformly $\mathbb{Q}$-integrable; by the Dunford-Pettis theorem, it is enough to prove that $K$ is relatively $\sigma(L^{1}(\mathbb{Q}),L^{\infty})$-compact. Let $\tau_{1}$ denote the subspace topology on $K$ with respect to $\tau$, and let $\tau_{2}$ denote the subspace topology on $K$ with respect to $\sigma(L^{1}(\mathbb{Q}),L^{\infty})$. By solidity of $\mathcal{Y}$,
\begin{equation*}
    \left\{\frac{d\mathbb{Q}}{d\mathbb{P}}Y:Y\in L^{\infty}(\mathbb{P})\right\}\subseteq\mathcal{Y},
\end{equation*}
and so the identity $(K,\tau_{1})\longrightarrow(K,\tau_{2})$ is continuous. Since a continuous image of a compact space is compact, $(K,\tau_{2})$ is compact, showing that $K$ is relatively $\sigma(L^{1}(\mathbb{Q}),L^{\infty})$-compact.
\end{proof}
\begin{lemma}\label{lem:coincide}
    Let $\mathbb{Q}$ denote the probability measure constructed in Lemma \ref{lem:weakcompacttopo} for a $K$-equicontinuous topology $\tau$. Then, if $C\subseteq\lambda K$ for some $\lambda>0$, the $\tau$ and $\sigma(L^{1}(\mathbb{Q}),L^{\infty})$ subspace topologies on $C$ are the same.
\end{lemma}
\begin{proof}
    Without loss of generality, we may take $\lambda=1$. Continuing with the notation from the proof of Lemma \ref{lem:weakcompacttopo}, the identity $(K,\tau_{1})\longrightarrow(K,\tau_{2})$ is continuous and bijective. But $(K,\tau_{1})$ is compact and $(K,\tau_{2})$ is Hausdorff, and hence $(K,\tau_{1})\longrightarrow(K,\tau_{2})$ is a homeomorphism.
\end{proof}
\subsection{Proof of Theorem \ref{thm:semicont}}\label{subsec:prf}
\begin{proof}[Proof of Theorem \ref{thm:semicont}]
    Let $\mathbb{Q}\sim\mathbb{P}$ be the probability measure constructed in Lemma \ref{lem:weakcompacttopo} for the $K$-equicontinuous topology $\tau$.
    \par
    We will first show that (\ref{it:lsc}) implies (\ref{it:fatou}). Fix an arbitrary $\lambda\in\mathbb{R}$ and $\lambda'\geq0$. Suppose that $(X^{n})_{n=1}^{\infty}\subseteq\{\varrho\leq\lambda\}\cap \lambda' K$, and $(X^{n})_{n=1}^{\infty}$ converges to $X\in\mathcal{X}$ in probability. We must show that $X\in\{\varrho\leq\lambda\}\cap \lambda'K$. By Vitali's convergence theorem and uniform $\mathbb{Q}$-integrability of $K$, $(X^{n})_{n=1}^{\infty}$ converges to $X$ in $L^{1}(\mathbb{Q})$-norm, hence also in $\sigma(L^{1}(\mathbb{Q}),L^{\infty})$. Lemma \ref{lem:coincide} thus implies $(X^{n})_{n=1}^{\infty}$ converges to $X$ in $\tau$. But by assumption, $\{\varrho\leq\lambda\}\cap\lambda'K$ is $\tau$-closed, and hence $X\in\{\varrho\leq\lambda\}\cap\lambda' K$, as desired.
    \par
    We now show (\ref{it:fatou}) implies (\ref{it:lsc}). Fix an arbitrary $\lambda\in\mathbb{R}$ and $\lambda'\geq0$. It suffices to show that $\{\varrho\leq\lambda\}\cap \lambda' K$ is $\tau$-closed. By Lemma \ref{lem:coincide}, it suffices to show that $\{\varrho\leq\lambda\}\cap \lambda' K$ is closed in $\sigma(L^{1}(\mathbb{Q}),L^{\infty})$. By the Hahn-Banach theorem and convexity of $\{\varrho\leq\lambda\}\cap\lambda'K$, this is equivalent to $L^{1}(\mathbb{Q})$-closedness of $\{\varrho\leq\lambda\}\cap \lambda' K$. But, as $\varrho$ has the $K$-Fatou property, $\{\varrho\leq\lambda\}\cap \lambda' K$ is closed in probability, hence closed in $L^{1}(\mathbb{Q})$, as desired.
\end{proof}
\section{Dual Representations}\label{sec:dualrep}
The equivalence between lower $\tau$-semicontinuity and the Fatou property established by and under the assumptions of Corollary \ref{corr:semicont-ks} can be related to the existence of dual representations.
\begin{theorem}\label{thm:dual}
     Let $\varrho:\mathcal{X}\longrightarrow\mathbb{R}\cup\{\infty\}$ be a convex and proper functional, and let $\tau=\sigma(\mathcal{X},\mathcal{Y})$ be a $K$-equicontinuous topology with the Krein-Šmulian property. Then the following are equivalent:
     \begin{enumerate}
        \item\label{it:fatou-dual} $\varrho$ has the $K$-Fatou property.
         \item\label{it:has-dual} $\varrho$ admits the dual representation
         \begin{equation*}
             \varrho(X)=\sup_{Y\in\mathcal{Y}}\left\{\mathbb{E}^{\mathbb{P}}(XY)-\varrho^{\ast}(Y)\right\}
         \end{equation*}
         for any $X\in\mathcal{X}$, where $\varrho^{\ast}(Y)=\sup_{Z\in\mathcal{X}}\left\{\mathbb{E}^{\mathbb{P}}(ZY)-\varrho(Z)\right\}$ for any $Y\in\mathcal{Y}$.
         \item\label{it:dual-lsc} $\varrho$ is lower semicontinuous with respect to $\tau$.
     \end{enumerate}
\end{theorem}
\begin{proof}
    (\ref{it:dual-lsc})$\iff$(\ref{it:has-dual}): this is a consequence of the Fenchel-Moreau theorem, noting that $\mathcal{Y}$ is identifiable with the continuous dual space of $(\mathcal{X},\tau)$.
    \par
    (\ref{it:dual-lsc})$\iff$(\ref{it:fatou-dual}): this is the content of Theorem \ref{thm:semicont}.
\end{proof}
Risk measures are usually considered to be monotone, and in this case it is possible to say more about the dual representation: the supremum may be taken over the positive or negative orthant in $\mathcal{Y}$.
\begin{corollary}\label{corr:positive-dual-rep}
    Let $\varrho:\mathcal{X}\longrightarrow\mathbb{R}\cup\{\infty\}$ be an increasing convex and proper functional, and let $\tau=\sigma(\mathcal{X},\mathcal{Y})$ be a $K$-equicontinuous topology with the Krein-Šmulian property. If $\varrho$ has the $K$-Fatou property, then $\varrho$ admits the dual representation
    \begin{equation*}
        \varrho(X)=\sup_{Y\in\mathcal{Y}_{+}}\left\{\mathbb{E}^{\mathbb{P}}(XY)-\varrho^{\ast}(Y)\right\}
    \end{equation*}
    for any $X\in\mathcal{X}$, where $\varrho^{\ast}(Y)=\sup_{Z\in\mathcal{X}}\left\{\mathbb{E}^{\mathbb{P}}(ZY)-\varrho(Z)\right\}$ for any $Y\in\mathcal{Y}$, and $\mathcal{Y}_{+}=\mathcal{Y}\cap L^{0}_{+}(\mathbb{P})$.
\end{corollary}
\begin{proof}
    The proof is standard, but included for the convenience of the reader.
    \par
    By Theorem \ref{thm:dual}, it suffices to show that, if $\varrho^{\ast}(Y)<\infty$, then $Y\geq0$. If not, putting $Z=\mathbf{1}_{\{Y<0\}}$, we have that $\mathbb{E}^{\mathbb{P}}(ZY)<0$. Fixing any $X\in\mathrm{dom}(\varrho)$ and applying Fenchel's inequality, we have that
    \begin{equation}\label{eq:fenchel-ineq-inc}
        \varrho(X)+\varrho^{\ast}(Y)\geq\varrho(X-\lambda Z)+\varrho^{\ast}(Y)\geq\mathbb{E}^{\mathbb{P}}((X-\lambda Z)Y)\geq\mathbb{E}^{\mathbb{P}}(XY)-\lambda\mathbb{E}^{\mathbb{P}}(ZY)
    \end{equation}
    for any scaling parameter $\lambda>0$. But while the left side of (\ref{eq:fenchel-ineq-inc}) is finite and bounded from below by $\mathbb{E}^{\mathbb{P}}(XY)-\lambda\mathbb{E}^{\mathbb{P}}(ZY)$, $\lambda$ and hence also $\mathbb{E}^{\mathbb{P}}(XY)-\lambda\mathbb{E}^{\mathbb{P}}(ZY)$ can be made arbitrarily large, a contradiction.
\end{proof}
Theorem \ref{thm:dual} requires the Krein-Šmulian property, which is automatically satisfied if $\mathcal{X}$ is the Banach space dual of $\mathcal{Y}$, but otherwise can be difficult to establish. However, if a risk measure $\varrho$ is only considered on some subset $C\subseteq\mathcal{X}$, which often corresponds to imposing constraints on an optimization problem, it is possible to establish a dual representation result without the Krein-Šmulian property.
\begin{theorem}\label{thm:local-dual-rep}
    Let $\varrho:\mathcal{X}\longrightarrow\mathbb{R}\cup\{\infty\}$ be a a convex and proper functional, and let $\tau=\sigma(\mathcal{X},\mathcal{Y})$ be a $K$-equicontinuous topology. If $\varrho$ has the $K$-Fatou property, and $C\subseteq\mathcal{X}$ is a non-empty $K$-bounded convex subset of $\mathcal{X}$ that is $\tau$-closed, then there exists a proper functional $\alpha:\mathcal{Y}\longrightarrow\mathbb{R}\cup\{\infty\}$ with
    \begin{equation*}
        \varrho(X)=\sup_{Y\in\mathcal{Y}}\left\{\mathbb{E}^{\mathbb{P}}(XY)-\alpha(Y)\right\}
    \end{equation*}
    for each $X\in C$.
\end{theorem}
\begin{proof}
    Put $\widetilde{\varrho}=\varrho+\mathbb{I}_{C}$, which inherits the $K$-Fatou property and convexity from $\varrho$ and $\mathbb{I}_{C}$. Since $C$ is $K$-bounded, we may find $\lambda'\geq0$ with $C\cap \lambda'K=C$. Fixing $\lambda\in\mathbb{R}$, we have that
    \begin{equation*}
        \left\{\widetilde{\varrho}\leq\lambda\right\}=\{\varrho\leq\lambda\}\cap C
    \end{equation*}
    and so $\left\{\widetilde{\varrho}\leq\lambda\right\}=\left\{\widetilde{\varrho}\leq\lambda\right\}\cap\lambda'K$. Since $\widetilde{\varrho}$ has the $K$-Fatou property, Theorem \ref{thm:semicont} implies $\left\{\widetilde{\varrho}\leq\lambda\right\}\cap\lambda' K$ is $\tau$-closed. So $\left\{\widetilde{\varrho}\leq\lambda\right\}$ is $\tau$-closed, and the arbitrariness of $\lambda\in\mathbb{R}$ implies $\widetilde{\varrho}$ is lower $\tau$-semicontinuous.
    \par
    By the same argument as the equivalence (\ref{it:dual-lsc})$\iff$(\ref{it:has-dual}) from Theorem \ref{thm:dual}, $\widetilde{\varrho}$ has the dual representation
    \begin{equation*}
        \widetilde{\varrho}(X)=\sup_{Y\in\mathcal{Y}}\left\{\mathbb{E}^{\mathbb{P}}(XY)-\alpha(Y)\right\},
    \end{equation*}
    for some proper functional $\alpha:\mathcal{Y}\longrightarrow\mathbb{R}\cup\{\infty\}$ for all $X\in\mathcal{X}$, and in particular for all $X\in C$. But $\varrho$ and $\widetilde{\varrho}$ coincide on $C$, so this proves the claim.
\end{proof}
Theorem \ref{thm:local-dual-rep} requires $\tau$-closedness of the constraint set $C$, but this is often harder to verify than other conditions, like closedness in probability. Thus, we note a version of Theorem \ref{thm:local-dual-rep} for sets which are closed in probability.
\begin{corollary}
    Let $\varrho:\mathcal{X}\longrightarrow\mathbb{R}$ be a a convex and proper functional, and let $\tau=\sigma(\mathcal{X},\mathcal{Y})$ be a $K$-equicontinuous topology. If $\varrho$ has the $K$-Fatou property, and $C\subseteq\mathcal{X}$ is a non-empty $K$-bounded convex subset of $\mathcal{X}$ that is closed in probability, then there exists a proper functional $\alpha:\mathcal{Y}\longrightarrow\mathbb{R}\cup\{\infty\}$ with
    \begin{equation*}
        \varrho(X)=\sup_{Y\in\mathcal{Y}}\left\{\mathbb{E}^{\mathbb{P}}(XY)-\alpha(Y)\right\}
    \end{equation*}
    for each $X\in C$.
\end{corollary}
\begin{proof}
    By Theorem \ref{thm:local-dual-rep}, it suffices to show $C$ is $\tau$-closed. Let $\mathbb{Q}\sim\mathbb{P}$ be from Lemma \ref{lem:weakcompacttopo}. Lemma \ref{lem:coincide} implies it suffices to show $C$ is $\sigma(L^{1}(\mathbb{Q}),L^{\infty})$-closed. As $C$ is convex, this is equivalent, by the Hahn-Banach theorem, to closedness in the norm of $L^{1}(\mathbb{Q})$. But $C$ is closed in probability, and hence also closed in the norm of $L^{1}(\mathbb{Q})$, as desired.
\end{proof}
\section{Acknowledgments}
The author would like to thank the two anonymous reviewers, whose comments have helped improve the article.
\printbibliography
\appendix
\section{Some Counterexamples}
\subsection{A Space with Trivial Dual}\label{app:trivial-dual}
Let $\mathcal{X}$ and $K$ satisfy the requirements of \S\ref{subsec:bounded}. In general, although $K$ is required to be closed and bounded in probability, this need not imply the price space $\mathcal{X}'_{K}=\mathrm{span}(K^{\circ})$ is non-trivial.
\begin{example}[$\mathcal{X}\neq\{0\}$ but $\mathcal{X}'_{K}=\{0\}$]
    Let $(X^{n})_{n=1}^{\infty}$ be an iid sequence of Cauchy-distributed random variables with scale parameter $1$ and location parameter $0$. Let $L$ be the absolutely convex hull of $(X^{n})_{n=1}^{\infty}$ and let $K$ be the $L^{0}(\mathbb{P})$-closure of $L$. For $\mathcal{X}=\mathrm{span}(K)$ to satisfy the requirements of \S\ref{subsec:bounded}, it is necessary and sufficient that $K$ is absolutely convex, closed in probability, and bounded in probability—only the last of which is not immediate.
    \par
    Being the closure of $L$ in probability, boundedness of $K$ is equivalent to boundedness of $L$. A generic $X\in L$ takes the form $X=\sum_{i}a_{i}X_{i}$ for a sequence $(a_{i})_{i}$ with finite support and $\Vert{a}\Vert_{\ell_{1}}\leq1$. Such an $X$ is Cauchy-distributed, with scale parameter $\Vert{a}\Vert_{\ell_{1}}$ and location parameter $0$, unless $\Vert{a}\Vert_{\ell_{1}}=0$, in which case $X=0$. Letting $\overline{F_{\gamma}}$ denote the complementary cumulative distribution function of a Cauchy distribution with scale parameter $\gamma>0$ and location parameter $0$, it therefore suffices to show
    \begin{equation*}
        \lim_{x\to\infty}\sup_{\gamma\in(0,1]}\overline{F_{\gamma}}(x)=0.
    \end{equation*}
    But $\frac{\partial}{\partial\gamma}\overline{F_{\gamma}}(x)=\frac{x}{\pi(x^{2}+\gamma^{2})}>0$ for $x>0$, so
    \begin{equation*}
        \lim_{x\to\infty}\sup_{\gamma\in(0,1]}\overline{F_{\gamma}}(x)=\lim_{x\to\infty}\overline{F_{1}}(x)=0
    \end{equation*}
    as desired.
    \par
    We now show triviality of the price space, namely, $\mathcal{X}'_{K}=\{0\}$. Suppose $\mathcal{X}'_{K}\neq\{0\}$, in which case there exists a nonnegative $Y\in K^{\circ}$ with $\mathbb{P}(\{Y>0\})>0$. Hence there is some $\varepsilon>0$ so the event $A=\{Y\geq\varepsilon\}$ satisfies $\mathbb{P}(A)>0$. We have that
    \begin{equation*}
        \mathbb{E}^{\mathbb{P}}\left(\vert{X}\vert\mathbf{1}_{A}\right)\leq\frac{1}{\varepsilon}
    \end{equation*}
    for all $X\in K$. Hence the sequence $\left(X^{n}\mathbf{1}_{A}\right)_{n=1}^{\infty}$ is bounded in $L^{1}(\mathbb{P})$. Komlós's theorem implies the existence of a strictly increasing sequence $(n_{m})_{m=1}^{\infty}\subseteq\mathbb{N}$ and $Y^{m}\in\mathrm{co}\left(\left\{X^{n_{m}+1},\dots,X^{n_{m+1}}\right\}\right)$ so that $(Y^{m})_{m=1}^{\infty}$ $\mathbb{P}$-a.s. converges to a finite random variable on $A$ (cf. the appendix of \cite{del-sch}). For $m\in\mathbb{N}$, let $\mathscr{G}_{m}=\sigma(Y^{m})$; then the event $B=\left\{\lim_{m\to\infty}Y^{m}\textrm{ exists and is finite}\right\}$ is an element of the tail $\sigma$-algebra $\bigcap_{m=1}^{\infty}\sigma\left(\bigcup_{k=m}^{\infty}\mathscr{G}_{k}\right)$ of the $\mathscr{G}_{m}$'s. As the $\mathscr{G}_{m}$'s are independent of each other, Kolmogorov's zero-one law applies to show $\mathbb{P}(B)\in\{0,1\}$. As $A\subseteq B$ and $\mathbb{P}(A)>0$, it thus follows that $\mathbb{P}(B)=1$. However, this contradicts the fact that a non-constant iid sequence cannot $\mathbb{P}$-a.s. converge. Hence $\mathcal{X}'_{K}=\{0\}$, as desired.
    \qed
\end{example}
\subsection{Necessity of Compactness}\label{app:necessary-compact}
Theorem \ref{thm:semicont} requires $K$ to be $\tau$-compact, at least for the implication (\ref{it:lsc})$\implies$(\ref{it:fatou}).
\begin{example}
    Let $\mathbb{P}$ be non-atomic, and define $\mathcal{X}=L^{1}(\mathbb{P})$, $K=B_{L^{1}}$, and $\mathcal{Y}=L^{\infty}(\mathbb{P})$. The topology $\tau=\sigma(\mathcal{X},\mathcal{Y})$ satisfies all the requirements of $K$-equicontinuity, except that $K$ is not $\tau$-compact. Despite the closeness of $\tau$ to being $K$-equicontinuous, Theorem \ref{thm:semicont} fails for $\tau$: $\varrho=\mathbb{E}^{\mathbb{P}}$ is such that $\{\varrho\leq\lambda\}\cap\lambda' K$ is $\tau$-closed for all $\lambda\in\mathbb{R}$ and $\lambda'\geq0$, but $\varrho$ does not have the $K$-Fatou property. Indeed, as $\mathbb{P}$ is non-atomic, there is a sequence of sets $(A_{n})_{n=1}^{\infty}\subseteq\mathscr{F}$ with $\mathbb{P}(A_{n})=\frac{1}{n}$, in which case the sequence $(X^{n})_{n=1}^{\infty}$, defined as $X^{n}=-n\mathbf{1}_{A_{n}}$, is $K$-bounded and $\mathbb{P}$-a.s. convergent to zero, but $\varrho(X^{n})=-1$ and $\varrho(0)=0$, so that $\liminf_{n\to\infty}\varrho(X^{n})<\varrho(0)$.
    \qed
\end{example}
\end{document}